# Experimental Proof of the Existence of a Bifurcation Process During the undrained test in Clay

## P. Evesque[1] & M. Hattab[2]

[1] Lab MSSMat, UMR 8579 CNRS, Ecole Centrale Paris
92295 Châtenay-Malabry, France, e-mail evesque@mssmat.ecp.fr
[2] LPMM/ISGMP, UMR 7554 CNRS, Université de Metz, île de Saulcy,
57045 METZ cedex 01, France, hattab@lpmm.univ-metz.fr

**Abstract:**
*Recent papers, based on a new simple incremental modelling which assumes an "isotropic" response, predicts that the trajectory followed during an undrained compression test exhibits a bifurcation process when the stress field arrives at q=M'p', i.e. indicating a discontinuous change of solution when arriving at q=M'p'. This paper looks at experimental data on dense sample obtained at p'=$c^{ste}$, and it shows that the trajectory (p'=$c^{ste}$, v=$c^{ste}$) does continue to exist at and beyond the q=M'p' plane. So, this demonstrates the validity of the analysis which uses the bifurcation theory and this strengthens the proposed modelling. Indeed, this demonstrates the reality of the bifurcation process during undrained compression.*

**Pacs # :** 5.40 ; 45.70 ; 62.20 ; 83.70.Fn

______________________________________________________________________

Recent papers [1-3] have proposed a new simple modelling to model the mechanical behaviour of granular media. It assumes an "isotropic" response [4] governed by two plastic parameters, *i.e.* the pseudo Young modulus $1/C_o$ and pseudo Poisson coefficient $\nu$.

$$\begin{pmatrix} d e_1 \\ d e_2 \\ d e_3 \end{pmatrix} = -C_o \begin{pmatrix} 1 & -\nu & -\nu \\ -\nu & 1 & -\nu \\ -\nu & -\nu & 1 \end{pmatrix} \begin{pmatrix} d s_1 \\ d s_2 \\ d s_3 \end{pmatrix} \quad (1)$$

As granular media exhibits hysteretis behaviour, the couple ($C_o,\nu$) shall be different for compression and extension. However it is assumed that a unique couple ($C_o,\nu$) is able to describe all the compression tests for a given sample under a given stress field $\sigma'$; of course, $C_o$ and $\nu$ do evolve with the stress and the strain and are fitted from experimental data from $\sigma'_2=\sigma'_3=c^{ste}$ compression test.

As $\sigma'_2=\sigma'_3=c^{ste}$ compression test obeys the Rowe's relation, one finds that $\nu$ shall depend on the stress ratio q/p' only according to Eq. (2), so that integration can be performed for different kind of fixed-strain compression [2-3]. Labelling M and M' respectively the following stress ratios of the critical state, *i.e.* ($\sigma'_1-\sigma'_3$)/$\sigma'_3$=M and M'=3($\sigma'_1-\sigma'_3$)/($\sigma'_1+\sigma'_2+\sigma'_3$)=q/p', Rowe'relation writes:

$$\nu = \sigma'_1/[2\sigma'_3(1+M)] \tag{2}$$

In particular trajectories of undrained tests can be found within this model. One finds in the (p',q) plane, which is the only one of interest in this special $\nu=c^{ste}$ test case, that the trajectory shall start vertically, *i.e.* q increases at constant p'; then it shall turn right (or left) on the q=M'p' line when the sample is dense (or loose) ; states such as q=M'p' are called characteristic states and exhibits a $\nu=1/2$. The turn on the q=M'p' line is due the conjunction of two facts: Firstly, because a new solution occurs at the point q=M'p' since $\delta\nu=0$ under any $\delta\sigma$ stress increment because $\nu=1/2$ when q=M'p'; so the trajectory q=M'p' is a possible trajectory. Secondly, because the solution q=M'p' is chosen effectively because it dissipates less energy than the solution $p'=c^{ste}$.

From an experimental point of view, the abrupt angular right turn is generally observed for dense samples, which reinforces the modelling. However, the trajectory for loose samples, consisting of an abrupt left turn, is not observed and much smoother right turn is observed instead. The interpretation of this last phenomenon is that the descent on the q=M'p' line corresponds to an unstable trajectory since "the evolution delivers always more energy than the sample dissipates", when taking second order terms. So, the material evolves spontaneously on the left prior having reached the q=M'p' line, for loose samples only; this smooth turn can be a consequence of induced anisotropy or on inhomogeneous behaviours.

Anyhow, this is not the point that we want to discuss here. Let us now discuss the angular turn on right, which is observed for dense samples. Indeed, the simple "isotropic" incremental modelling attributes this turn to a bifurcation process. It would be important to demonstrate experimentally the existence of this bifurcation.

One way to prove the existence of this bifurcation consists in demonstrating that at q=M'p' location the system has the possibility to chose in between the two behaviours during an undrained compression, *i.e.* trajectory (i) (q=M'p', $\nu=c^{ste}$) trajectory *vs.* trajectory (ii) ($p'=c^{ste}$, $\nu=c^{ste}$). First possibility is obviously demonstrated, since it is the one that is observed truly. In order to prove the second possibility, one can perform a compression test at $p'=c^{ste}$ on dense samples. In this case, the "isotropic" incremental modelling predicts that $\delta\nu=0$ whatever q, as far as the sample response remain "isotropic" [4]. So, if one observes during such a compression test that the volume variation remains equal to 0 when the trajectory cuts the q=M'p' line, this means that an undrained compression can also follow the line $p'=c^{ste}$ when crossing this line. This will then indicate that both trajectories dq=M'dp' and dp'=0 are possible on the location defined by q=M'p'.

Indeed this is just what is observed in experimental data reported in Fig. 1. It concerns compression tests on Kaolinite clay at $p'=c^{ste}$ on over-consolidated clay samples [5] at different degree of over-consolidation. In this Fig 1, all tests start at constant volume, which validates the isotropic incremental modelling of Eq. (1); furthermore, Fig. 1 demonstrates that this isotropic incremental modelling remains valid beyond the q=M'p' line for large enough over-consolidation. This demonstrates that trajectory (ii) remains possible for undrained tests on dense sample and demonstrate that real undrained test exhibits a pure bifurcation process when reaching the q=M'p' line.

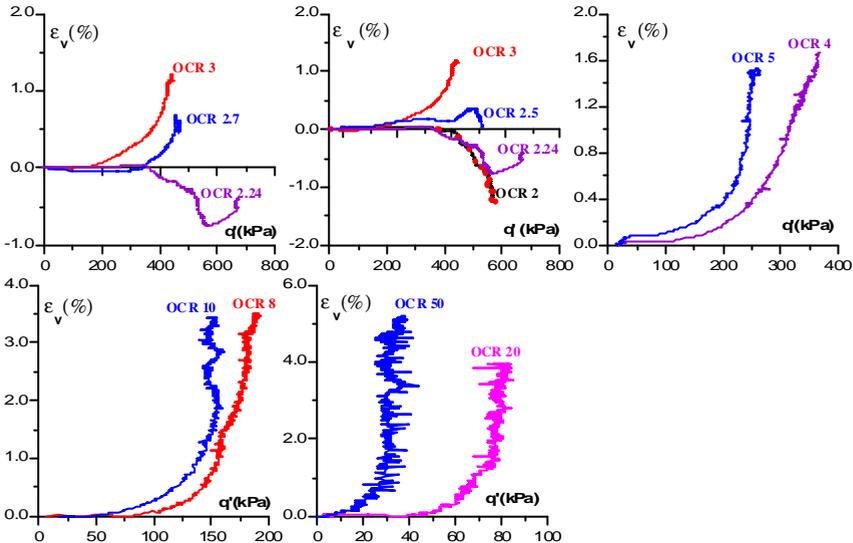

**Figure 1:** *compression test at mean constant pressure* $p'=(\sigma'_1+\sigma'_2+\sigma'_3)/3=c^{ste}$ *, performed on Kaolinite clay: tests on over-consolidated clays (OCR>2) demonstrate that volume remains constant even above the* q=M'p' *line. This demonstrates that the trajectory of an undrained test could stay on the* $p'=c^{ste}$ *line if bifurcation was not occuring spontaneously.*

**Conclusion:**

This paper shows that two trajectories can be followed at the same time during an undrained test when q=M'p', *i.e.* the one which is characterised by $p'=c^{ste}$ and the one characterised by q=M'p'. The one which is chosen spontaneously corresponds to q=M'p'; this is due most likely to energetic consideration as proposed in [1-3]. So, starting from p'=0, the real trajectory followed by an undrained test shall follows the trajectory $p'=c^{ste}$, then it bifurcates on the path q=M'p', when it arrives at this line. This proves the validity of the approach proposed in [1-3] and reinforce the validity of the concepts that are developed there.

Indeed, this paper demonstrates also the validity of the concept of the characteristic states, which are those states with $\nu=1/2$. These states have not to be confused with the critical states, since they do exhibit a non-infinite $C_o$ which is not the case for the critical states. In fact the space made of all the critical state is a curve in the (q,p',v) space and it is a subspace of the space made of all the characteristic states, which is a surface in the (q,p',v) space.

As both trajectories, *i.e.* (i) (q=M'p',v=$c^{ste}$) and (ii) (p'=$c^{ste}$, v=$c^{ste}$), remain possible when q reaches the value M'p' during the undrained tests, this explains why one can observe some fluctuation on the experimental behaviour at that peculiar point, and why also advice diverges depending on the specialist.

*Acknowledgements:* The authors want to thank Prof. J. Biarez for helpful discussion and P.E. thanks CNES for partial funding.

The electronic arXiv.org version of this paper has been settled during a stay at the Kavli Institute of Theoretical Physics of the University of California at Santa Barbara (KITP-UCSB), in june 2005, supported in part by the National Science Fundation under Grant n° PHY99-07949.

*Poudres & Grains* can be found at :
http://www.mssmat.ecp.fr/rubrique.php3?id_rubrique=402